\documentstyle[aaspp4,psfig]{article}

\begin{document}
\title{Observational Tests of the Mass-Temperature Relation 
  for Galaxy Clusters}

\author{Donald J. Horner\altaffilmark{1}, Richard F. Mushotzky} 
\affil{Laboratory for High Energy Astrophysics, NASA/GSFC, Code 662,
  Greenbelt, MD 20771}
\and 
\author{Caleb A. Scharf} \affil{Space Telescope Science Institute, 3700
  San Martin Drive, Baltimore, MD 21218}
\altaffiltext{1}{Also Astronomy Department, University of Maryland}

\begin{abstract}
  We examine the relationship between the mass and x-ray gas
  temperature of galaxy clusters using data drawn from the literature.
  Simple theoretical arguments suggest that the mass of a cluster is
  related to the x-ray temperature as $M \propto T_{\rm x}^{3/2}$.
  Virial theorem mass estimates based on cluster galaxy velocity
  dispersions seem to be accurately described by this scaling with a
  normalization consistent with that predicted by the simulations of
  Evrard, Metzler, \& Navarro (1996).  X-ray mass estimates which
  employ spatially resolved temperature profiles also follow a $T_{\rm
    x}^{3/2}$ scaling although with a normalization about 40\% lower
  than that of the fit to the virial masses.  However, the isothermal
  $\beta$-model and x-ray surface brightness deprojection masses
  follow a steeper $\propto T_{\rm x}^{1.8-2.0}$ scaling.  The
  steepness of the isothermal estimates is due to their implicitly
  assumed dark matter density profile of $\rho(r) \propto r^{-2}$ at
  large radii while observations and simulations suggest that clusters
  follow steeper profiles (e.g., $\rho(r) \propto r^{-2.4}$).
\end{abstract}

\section{Introduction}\label{sec:intro}

The relationship between the mass and x-ray temperature of galaxy
clusters is a necessary bridge between theoretical Press-Schechter
models, which give the mass function (MF) of clusters, and the
observed x-ray temperature function (TF) (e.g., \cite{edge90};
\cite{ha91}; \cite{henry97}; \cite{markevitch98}).  Theoretical
arguments suggest that the virial mass of a galaxy cluster is simply
related to its x-ray temperature as $M \propto T_{\rm x}^{3/2}$.
These arguments are supported by simulations which show a tight
correlation between mass and temperature (e.g., \cite{emn96};
\cite{bn98}; \cite{enf98}).  This suggests that the M--T
relationship may also be a relatively accurate and easy way to
estimate cluster mass.  However, the M--T relationship first needs to
be calibrated using masses estimated by some other means.

The oldest method of measuring cluster mass is the virial mass
estimate based on dynamical analysis of the observed velocity
dispersion of the cluster galaxies.  The existence of the x-ray
emitting ICM of galaxy clusters allows an independent mass estimate
but requires knowledge or assumptions about both the x-ray temperature
and surface brightness profiles.  More recently, strong and weak
gravitational lensing by clusters has provided a third independent
mass estimate.  If clusters are dynamically relaxed and relatively
unaffected by non-gravitational process, these three methods should
give the same results.  In this paper, we concentrate on virial and
x-ray mass estimates.

In Section~\ref{sec:mxtx}, we discuss the theoretical basis of the
M--T relation and the results of cluster simulations.  We then compare
this relation with those using masses based on galaxy velocity
dispersions (Section~\ref{sec:dispersion}), x-ray mass estimates of
clusters with spatially resolved x-ray temperature profiles
(Section~\ref{sec:noniso}), and isothermal x-ray mass estimates
($\beta$-model estimates in Section~\ref{sec:beta} and surface
brightness deprojection in Section~\ref{sec:deproj}). In
Section~\ref{sec:conclusions}, we discuss the results and present
conclusions.


\section{Theory and Simulations}\label{sec:mxtx}

For gas that shock heats on collapse to the virial temperature of the
gravitational potential, the average x-ray temperature
\begin{equation}
  \label{eq:virialx}
  T_{\rm x} \propto \frac{M_{\rm vir}}{r_{\rm vir}} \propto M_{\rm
  vir}^{2/3} 
\end{equation}
where $r_{\rm vir}$ is the virial radius, the boundary separating the
material which is close to hydrostatic equilibrium from the matter
which is still infalling.  The coefficient of this relationship is a
complicated function of cosmological model and density profile of the
cluster (see e.g., \cite{lilje92}).  However, because the infall
occurs on a gravitational timescale $t_{\rm grav} \propto
\rho^{-1/2}$, the virial radius should occur at a fixed value of the
density contrast, defined as:
\begin{equation}
  \label{eq:over}
  \delta = \frac{\overline{\rho(r)}}{\rho_{c}(z)} = 
  \frac{M(<r)}{\frac{4}{3} \pi \rho_{c}(z)r^3}
\end{equation}
where $\rho_{c}(z)$ is the critical density.  For an $\Omega_{0} = 1$
universe, $\delta_{\rm vir} = 18 \pi^2 \approx 178$ but drops to lower
density contrasts for lower values of $\Omega_{0}$ (e.g.,
\cite{lc93}).  Since we do not have {\it a priori} knowledge of the
actual value of $\delta_{\rm vir}$, we scale all of our results to
$\delta$ = 200, which should contain only virialized material, and has
been used previously by other authors (e.g., \cite{cnoc96};
\cite{nfw95}).

Evrard et~al. (1996) (hereafter EMN) present a M--T relation which
seems to well describe simulated clusters in six different
cosmological models (two $\Omega_{0}=0.2$ and four $\Omega_{0}=1$
models, see their Table~1 for details).  EMN assume that $M \propto
T_{\rm x}^{3/2}$ and then fit the coefficient of the relationship at
various density contrasts.  For $\delta \approx 200$, the coefficient
depends only weakly on $\Omega_{0}$, with a difference of $\approx
5$\% between the $\Omega_{0} = 0.2$ and $\Omega_{0} = 1$ models while
the difference rises to $\approx 40$\% at $\delta = 2500$.

From EMN's Equation 9 and fitting the normalization using their
Table~5 (excluding the $\delta = 2500$ values but including both
$\Omega_{0} = 0.2$ and $\Omega_{0} = 1$ points), the expression for
mass as a function of temperature and density contrast is:
\begin{equation}\label{mdelta}
  M(\delta,T_{\rm x}) = (1.81 \pm 0.23) \times 10^{14} 
  \delta^{-0.266 \pm 0.022} 
   \left( \frac{T_{\rm x}}{{\rm keV}} \right)^{3/2} \
    h^{-1}\ {\rm M_{\odot}}
\end{equation}
where {\it h} is the Hubble constant in units of 100 km~s$^{-1}$
Mpc$^{-1}$.  Note that Equation~\ref{mdelta} shows that clusters in
the EMN simulations have dark matter density profiles $\rho(r) \propto
r^{-2.4}$ in accordance with the effective slope of the universal
density profile of \cite{nfw96} in the relevant range of radii.

Other simulations using different cosmological models and codes
generally give normalizations similar to Equation~\ref{mdelta} to
within $\lesssim 20$\%.  Eke et~al. (1998) give masses and gas
temperatures at a density contrast of 100 for their simulations of an
$\Omega_{0} = 0.3$ $\Omega_{\Lambda} = 0.7$ model which are well
described by Equation~\ref{mdelta}.  \cite{bn98} give the M--T at
$\delta = 250$ for simulations using variety of cosmological models.
Like EMN they find the normalization is fairly insensitive to the
model used although their normalizations are about 20\% higher.
Lastly, \cite{mte97} examine the differences between simulations of
galaxy clusters with and without galactic winds.  The M--T relation in
the wind models is slightly steeper due to the energy injection, with
a power law index $\sim 1.6$ and has temperatures about 20\% higher at
the low mass end.

A redshift dependence is introduced into the normalization of the M--T
relation by the definition of density contrast since the critical
density is a function of $\Omega$ and the redshift of formation (see
e.g., \cite{lilje92}; \cite{ecf96}; \cite{vd98}).  This should not
substantially affect our results as the samples considered consist
mainly of low redshift ($z \lesssim 0.1$) objects and/or have scatter
in the mass estimates considerably greater than the effect introduced
by the redshift dependence.

Since we have restricted ourselves to low density contrasts and
redshifts, we will not discuss gravitational lensing mass estimates.
Lensing estimates are usually limited to high density contrasts
$\delta \ge 3000$ (even for weak lensing) and to moderate-to-high
redshifts.  However, \cite{hov98} have reported good agreement
between the EMN relation and their sample of eight lensing clusters.
They assume $M \propto T_{\rm x}^{3/2}$ and an $\Omega_{0} = 1$
cosmology.  Their best fit normalization is 0.82 $\pm$ 0.38 (rms
dispersion) times the EMN normalization.


\section{Virial Theorem Mass Estimates}\label{sec:dispersion}

Assuming that the galaxies are distributed similarly to the total
mass, the virial mass of a cluster depends on the line of sight
projected velocity dispersion of the galaxies, $\sigma_{p}$, and the
virial radius, $r_{vir}$:
\begin{equation}\label{eq:virial}
  M_{vir} = \frac{3}{G} \sigma_{p}^2 r_{vir}.
\end{equation}

If the entire system is not included in the observational sample, as
is common for galaxy clusters, Equation~\ref{eq:virial} overestimates
the mass.  The usual form of the virial theorem (2U + T = 0) should be
modified to include the surface term (2U + T = 3PV) since the surface
pressure reduces the amount of mass needed to keep the system in
equilibrium (see \cite{girardi98b}; \cite{cnoc97}).

Girardi et~al. (1998) (hereafter G98) have derived virial masses for
170 nearby clusters ($z \leq 0.15$) using data compiled from the
literature and the ENACS data set \cite{katgert98}.  They define the
virial radius to be $r_{vir} = 0.002 \sigma_{p} \ h^{-1}$ Mpc where
$\sigma_{p}$ is in km s$^{-1}$ and consider only galaxies within this
radius in the mass estimation.  Their quoted masses are generally
smaller than previous estimates by $\lesssim 40\%$ which they
attribute to stronger rejection of interlopers and a correction factor
of $\sim 19\%$ accounting for the surface term.

We have cross-correlated the G98 catalog with two catalogs of ASCA
temperatures (\cite{fukazawa97}; \cite{markevitch98}) to obtain a
subsample of 30 clusters with at least 30 redshifts within $r_{vir}$.
Fukazawa and Markevitch excluded the center of the clusters to
minimize the effect of cooling flows on the derived temperature.  In
cases of multiple measurements, the temperatures generally agree
within their quoted errors (usually $\lesssim$ 10\%), and we have
preferentially used the Fukazawa temperatures.  This sample has few
low temperature clusters, so we have analyzed the archival ASCA data
of three additional G98 clusters (see Table~\ref{tab:clusters}).
Cooling flow effects are not important for these three clusters.
Table~\ref{tab:clusters} lists clusters in the final sample along with
their adopted x-ray temperature (column 2), G98 virial radius (column
3), and virial mass (column 4).

The assumption that $r_{vir} = 0.002 \sigma_{p}$ is quite approximate,
and the actual relation between $r_{vir}$ and $\sigma_{p}$ depends on
the cosmological model.  The density contrast of the G98 virial masses
($= M_{vir}/(4/3 \pi r_{vir}^{3} \rho_{c})$) is generally less than
200 with a mean (and standard deviation) of $97 \pm 23$.  Assuming the
dark matter density $\rho(r) \propto r^{-2.4}$ in the outer parts of
the clusters, which both the EMN simulations and Girardi data (at
least the galaxy distribution) seem to follow, we have rescaled their
masses to $\delta = 200$.  Effectively this is just a change of
normalization such that the rescaled masses are smaller than $M_{vir}$
by about 15\% (with a standard deviation of about 5\%).

\begin{figure}
  \psfig{figure=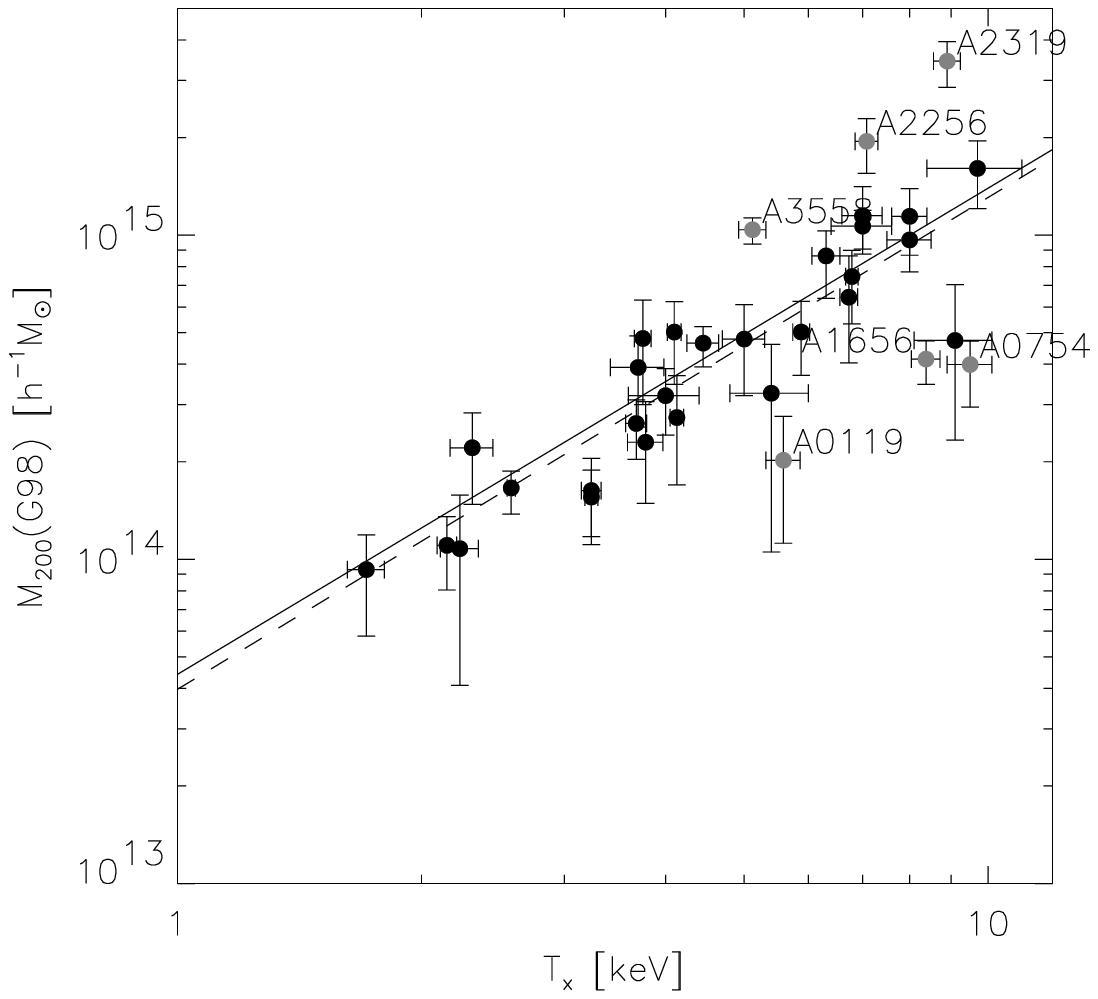}\caption{Virial Mass-Temperature relation.
    The solid line is the theoretical EMN relation while the dashed
    line is a fit to the G98 virial masses (rescaled to an density
    contrast of 200) and ASCA temperatures.  The error bars represent
    the 90\% confidence intervals for temperature and 68\% confidence
    for mass.  The gray circles indicate clusters excluded from the
    fit (named in the plot).
  \label{f1.eps}}
\end{figure}

Figure~\ref{f1.eps} shows the distribution of rescaled virial
masses versus temperature for this subsample.  A power law fit using
the BCES bisector method of \cite{ab96}, which takes into account the
errors in both variables and the possibility of intrinsic scatter,
gives $M \propto T_{\rm x}^{1.71 \pm 0.20}$ (all quoted errors are
1$\sigma$ unless otherwise stated), marginally inconsistent with the
EMN relation.  The seven most severe of the outliers in this plot are
A119, A754, A1656 (Coma), A2256, A2319, A3558, and A2029. Except for
A2029, all are known to contain complex velocity or temperature
structure.  We have marked these six clusters in gray in
Figure~\ref{f1.eps}.  Removing these six clusters from the fit
gives $M \propto T_{\rm x}^{1.53 \pm 0.13}$ and a normalization
consistent with the EMN relation (see Table~\ref{tab:results}).
Further outlier removal or permutations of the temperatures (i.e.\ 
using Markevitch instead of Fukazawa temperatures) does not cause
significant differences in the fit (e.g., power law index is changed
by $\pm 0.05$).  Given the relatively large scatter, more clusters
with well measured temperatures, especially cooler/less massive
clusters, are needed to further constrain the relationship between
virial mass and x-ray temperature.

\subsection{The Velocity Dispersion--Temperature Relation} \label{sec:sigt}

\begin{figure}
  \psfig{figure=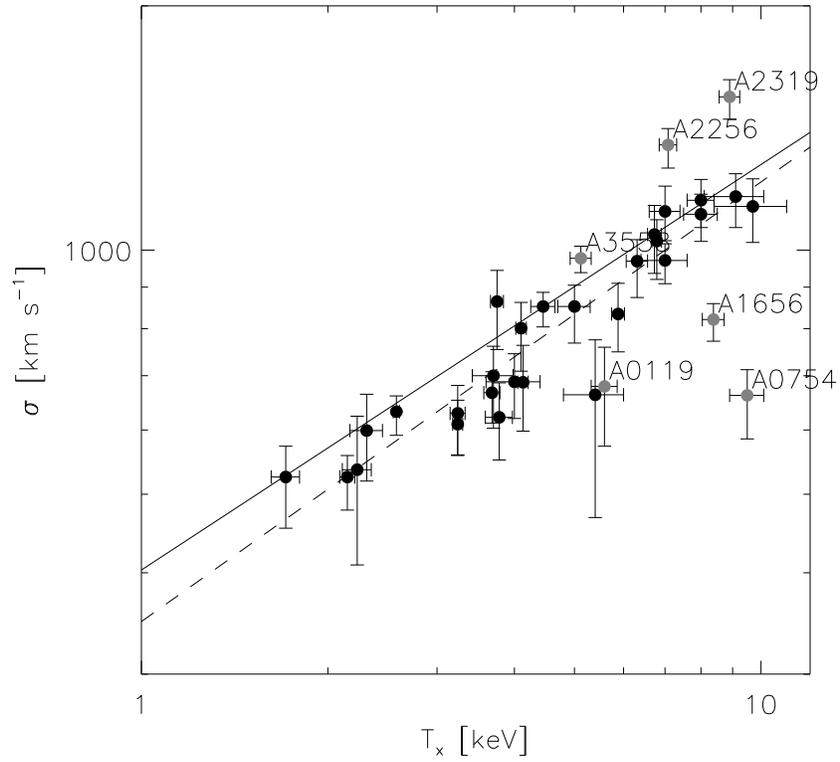}\caption{Velocity Dispersion-Temperature
    relation.  The solid line is $\sigma \propto T_{\rm x}^{1/2}$.
    The dashed line is a fit to the data excluding the clusters marked
    in gray.  The error bars represent the 90\% confidence intervals
    for temperature and 68\% confidence for velocity dispersion.
  \label{f2.eps}}
\end{figure}

To first order, $M_{vir} \propto \sigma^{3}$ since $M_{vir} \propto
\sigma^{2} r_{vir}$ and $r_{vir} \propto \sigma$.  Therefore, the
virial mass--temperature relation is related to the more extensively
discussed velocity dispersion--temperature ($\sigma$--$T_{\rm x}$)
relation (e.g., \cite{lb93}; \cite{bmm95}; \cite{wu98c} and
references therein).  Figure~\ref{f2.eps} shows the $\sigma$ --
$T_{\rm x}$ relation for our G98 subsample.  Excluding the same six
clusters as above as fit gives $\sigma \propto T_{\rm x}^{0.54 \pm
  0.03}$.  This is consistent with the observed $M_{vir} \propto
T_{\rm x}^{1.5 \pm 0.1}$.

If galaxies and gas are both in equilibrium with the cluster potential
and gravity is the only source of energy, $\sigma \propto T_{\rm
  x}^{1/2}$ (e.g., Bird et~al. 1995).  Our results are marginally
steeper than this and consistent with many previous estimates (e.g.,
0.61 $\pm$ 0.13 from Bird et~al. 1995).  However, our fit is shallower
than the fit G98 give in their paper (0.62 $\pm$ 0.04) possibly due to
improved x-ray temperatures since they use Einstein and Ginga
temperatures from \cite{david93} and \cite{wjf97}.  However, our fit
is definitely shallower than the 0.67 $\pm$ 0.09 found by Wu et~al.
(1998).  While they draw a much larger sample (94 clusters) from the
literature, their sample is heterogeneous and the quality of their
data is unclear.  In contrast, we are using only G98 derived velocity
dispersion with greater than 30 member redshifts and precise ASCA
temperatures.

\subsection{Scatter in Virial Mass Estimator} \label{sec:scatter}

The scatter in the virial mass estimator is expected to be quite large
because of shot noise due to the finite number of galaxies in a
cluster and projection effects due to contamination by background and
foreground galaxies.  The scatter in the observed virial
mass--temperature relation is then a combination of the dispersion in
the virial mass estimator (with respect to the true cluster mass) and
any intrinsic dispersion in the M--T relation.

\begin{figure}
  \psfig{figure=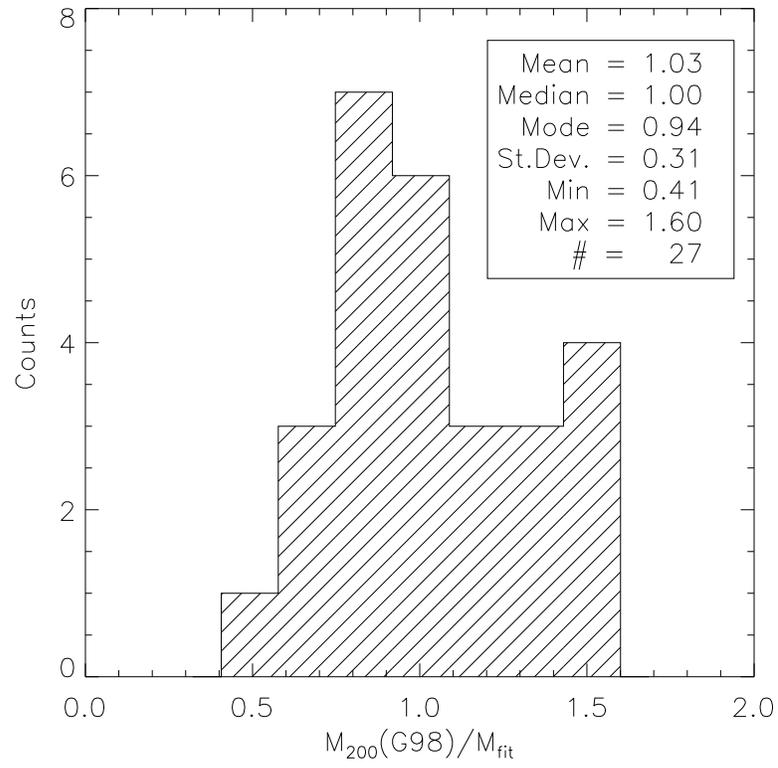}\caption{Histogram of virial to best fit
    masses excluding outliers.
  \label{f3.eps}}
\end{figure}

Figure~\ref{f3.eps} shows a histogram of the ratio of the
rescaled virial masses ($M_{200}(G98)$) to the mass expected from the
best fit relation ($M_{fit}$).  The mean (or median) is approximately
1.0 with a standard deviation of 0.31.  Figure~\ref{f3.eps}
excludes the six outliers that were not fitted, including these
clusters increases the standard deviation to 0.59.  The expected
scatter in the virial mass estimator has not been widely reported in
the literature, but \cite{fb84} find that in their simulations of
galaxy clusters the ratio of the virial mass to true mass is $0.97 \pm
0.36$ (1$\sigma$ standard deviation) after removing contaminating
background and foreground galaxies.  This predicted scatter is close
to the observed scatter around the fit and suggests the dispersion in
the virial mass -- temperature relationship is primarily due to the
scatter in the virial mass estimator.

\begin{figure}
  \psfig{figure=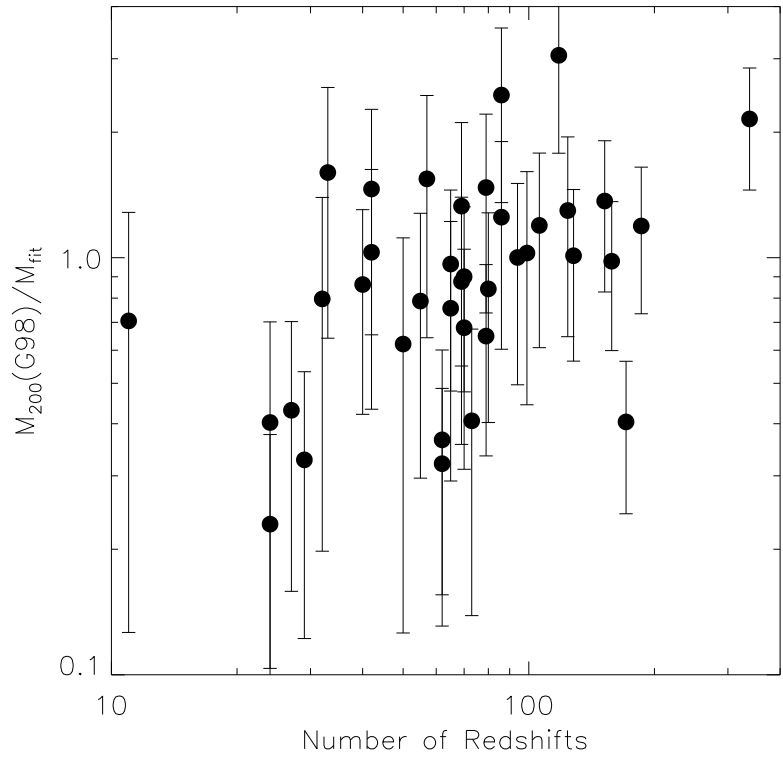}\caption{Ratio of virial to best fit
    masses as function of number of redshifts used to calculate virial
    mass.
  \label{f4.eps}}
\end{figure}

This is further supported by the distribution of
$M_{200}(G98)/M_{fit}$ as a function of the number of redshifts
($n_z$) used to calculate the virial mass (see
Figure~\ref{f4.eps} which also includes clusters with less than
30 redshifts).  The scatter is about a factor of 2 lower for clusters
with $n_z \ge 80$.  This is not surprising as a larger number of
redshifts increases the accuracy of the virial mass estimator by
decreasing the shot noise.  Together with the results in Figure 3,
this further indicates that the M-T relation must have very small
intrinsic scatter.


\section{X-ray Mass Estimates}\label{sec:xray}

X-ray mass estimates are based on the assumptions that the ICM is in
hydrostatic equilibrium and supported solely by thermal pressure.
With the further assumption of spherical symmetry, the gas density,
temperature, and pressure are related to the mass by:
\begin{equation}  \label{eq:deproj1}
  \frac{dP_g}{dr} = \rho_{g} \frac{GM(<r)}{r^2}  
\end{equation}
\begin{equation}  \label{eq:deproj2}
 P_g = \frac{\rho_g k T_x}{\mu m_p}.
\end{equation}
\noindent  where k is Boltzmann's constant, and $\mu m_{\rm p}$
is the mean molecular weight of the gas.  The enclosed mass at a
radius, r is then:
\begin{equation}
  \label{eq:hydro}
  M(<r) = - \frac{kT_{\rm x}(r)}{G \mu m_{\rm p}} r \left[ \frac{d\ log\ 
  \rho_{g}(r)}{d\ log\ r} + \frac{d \ log\ T_{\rm x}(r)}{d\ log\ r} \right],
\end{equation}
\noindent which depends on both the gas density and temperature
profiles.  The gas density is usually assumed to follow a $\beta$
profile: 
\begin{equation}
  \label{eq:bmodel}
  \rho_{g}(r) = \rho_{g}(0) \left[ 1 + \left(\frac{r}{r_{\rm c}} \right)^{2}
  \right]^{-3 \beta /2}.
\end{equation}

Historically, x-ray detectors have either had good spatial or spectral
resolution but not both.  Measuring the actual temperature profiles of
clusters has really only become practical with the advent of ASCA and
its ability to obtain spatially resolved spectra.  The combination of
using ROSAT to obtain $\rho_{g}(r)$ and ASCA to obtain $T_{\rm x}(r)$
can yield cluster masses with unprecedented accuracy.  However,
estimating the temperature profile is complicated by the spatial and
energy of the ASCA PSF, and mass estimates have only been reported for
a few of the best observed clusters. The difficulty of measuring the
temperature profiles means that there are far larger samples of
clusters for which only the average x-ray temperature and isothermal
mass estimates are available.

\subsection{Mass Estimates with Spatially Resolved Temperature Profiles}
\label{sec:noniso}

No large catalog of clusters with masses measured using spatially
resolved temperature profiles has been published.  Therefore, we have
searched the literature (including conference proceedings) to obtain a
sample of 12 clusters with masses measured using known temperature
profiles.  These clusters are presented in Table~\ref{tab:clusters}
with the adopted average x-ray temperature and 90\% confidence limits
(column 2 or table notes), largest radius in which the mass was given
(column 5), the mass within that radius (column 6), and the reference
from which the data was taken (column 10).  If given, we have used
global temperature values given by the respective authors. Otherwise,
we have taken them from catalogs of ASCA temperatures as in
Section~\ref{sec:dispersion}.  The formal errors on these mass
estimates are small as they are well constrained by the density and
temperature profiles.  However, the systematic uncertainties (i.e.\ 
uncertainties in the ASCA PSF and effective area) are much more
difficult to quantify. For fitting purposes, we have chosen not to
weight the fit with any mass errors, only with the errors in
temperature.

Several of the clusters used warrant some comments.  The masses for
A496 and A2199 from \cite{mlaf94} were derived without ASCA PSF
corrections.  However, the corrections are not large for these
clusters as they are relatively cool and only the central
field-of-view of the telescope was used.  A2256 is known merger, but
\cite{mv97} argue that the subcluster is physically well separated
along the line of sight and has not disturbed the bulk of the primary
cluster's gas.  Although this cluster was considered an outlier in the
virial mass fit, this can be attributed due to contamination of the
optical velocity dispersion measurement.

As with the virial masses, we have rescaled the masses to a density
contrast of 200 assuming the dark matter density profile $\rho(r)
\propto r^{-2.4}$.  On average, the effect of rescaling is to increase
the masses by an average of 20\% with a standard deviation of
$\approx$25\%.  There is some support for using this profile.
\cite{mmiy96} and \cite{mv97} report similar profiles for A2163 and
A2256.  While other authors in Table~\ref{tab:clusters} do not give
density profiles, some (e.g., \cite{ohashi97}; \cite{swm98})
contain plots of the mass as a function of radius which are 
consistent with the $\rho(r) \propto r^{-2.4}$ profile.  Lastly, the
clusters themselves seem to obey this profile.  Figure~\ref{f6.eps}
show the density contrast (effectively the average density) as a
function of radius.  We normalize the radius to EMN's $r_{200}$, the
radius at which the density contrast is 200, so that we are comparing
similar scales in different clusters but this makes little difference.
An unweighted fit indicates that $\rho(r) \propto r^{-2.4 \pm 0.1}$.

\begin{figure}
  \psfig{figure=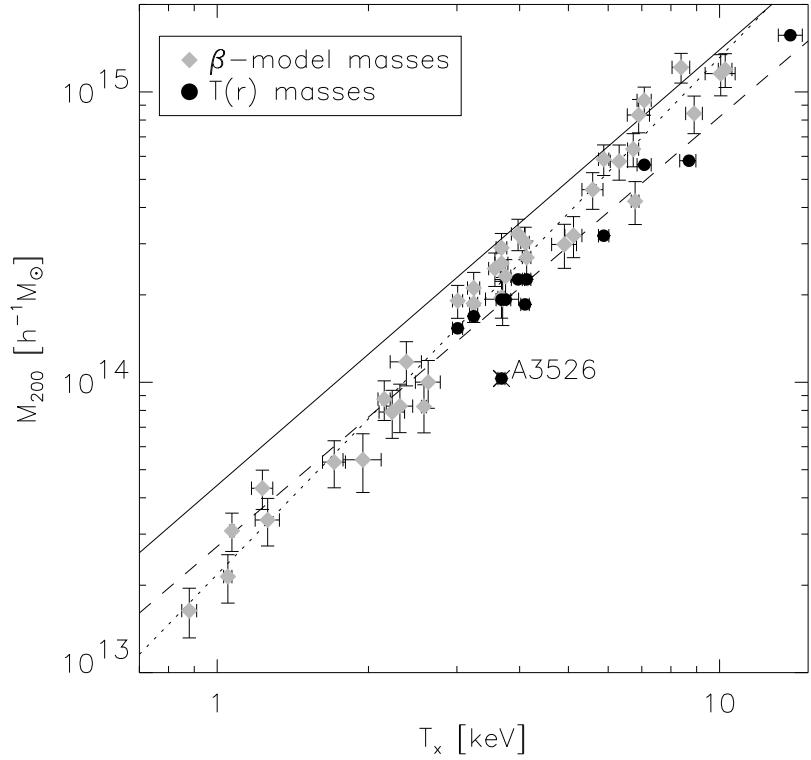}\caption{Mass--Temperature Relation.
    Filled circles are clusters with masses measured using spatially
    resolved temperature profiles rescaled to an density contrast of
    200.  The dashed line is a fit to this data.  Gray diamonds are
    isothermal $\beta$-model masses within an overdensity of 200 for
    the F97 sample.  The dotted line is the corresponding fit.  The
    solid line is the theoretical EMN relation.  Error bars are 90\%
    confidence limits.
  \label{f5.eps}}
\end{figure}

\begin{figure}
  \psfig{figure=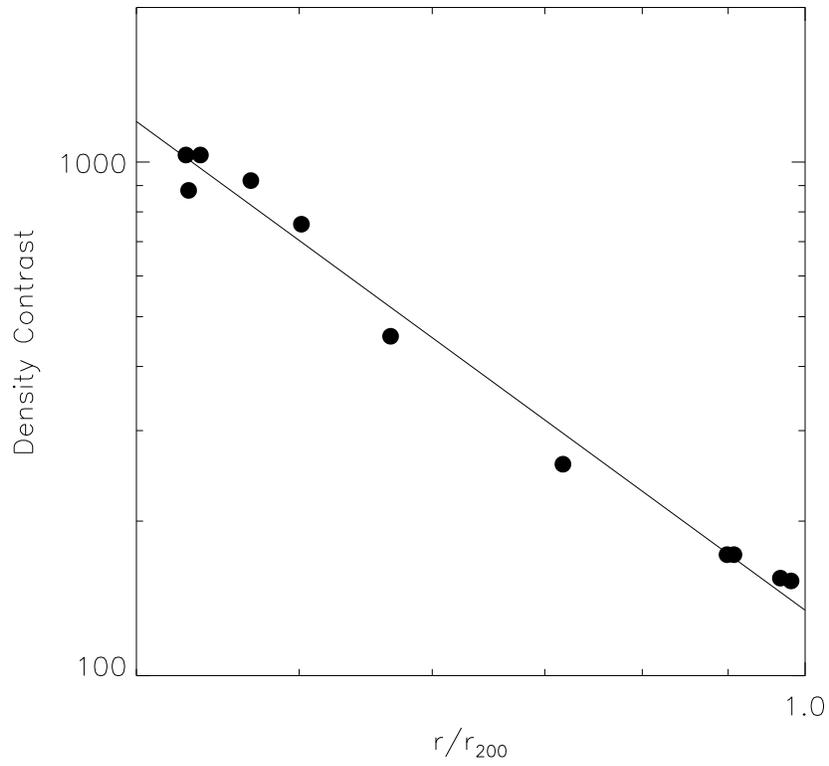}\caption{Density profiles of clusters with
    resolved temperature profiles as a function of radius, normalized
    to the EMN's $r_{200}$.  The solid line represents a fit to the
    data, $\delta \propto (r/r_{200})^{-2.4 \pm 0.1}$.
  \label{f6.eps}}
\end{figure}

In Figure~\ref{f5.eps}, we plot the rescaled masses versus
temperature.  Excluding A3526 (Centaurus), the best fit is $M \propto
T_{\rm x}^{1.48 \pm 0.12}$ but with a normalization about 40\% lower
than that of the EMN relation.  Given the heterogeneous nature of the
sample, the dispersion around this fit is surprisingly small
($\lesssim$ 10\% in mass) indicating that the intrinsic correlation
between temperature and mass is quite tight.  Using a different
density profile to extrapolate the mass to a density contrast of 200
has the tendency to increase the dispersion in this fit (i.e.\ about
25\% for $\rho(r) \propto r^{-2}$) but does not have much effect on
the power law index or normalization of the fit.

The lower normalization than that found using the EMN relation or
virial masses may reflect systematics in the masses derived using the
temperature profiles, or it could be a problem with the simulations
and systematics in the virial mass determinations.  \cite{ap98} have
found that the virial mass estimate can either overestimate or
underestimate the mass depending on the aperture of the region
sampled.  However, a direct comparison of the virial mass and x-ray
masses for the 9 clusters with temperature profiles masses
($M_{200}(T(r))$) and G98 virial masses ($M_{200}(G98)$)
(Figure~\ref{f8.eps}) shows no clear trend.  Further simulations and
future observations of clusters with temperature profiles and virial
masses will be need to explore this issue further.

\subsection{The Isothermal $\beta$-model}\label{sec:beta}

The most often used x-ray mass estimator has been the isothermal
$\beta$-model which assumes that the gas is isothermal and that the
gas density follows Equation~\ref{eq:bmodel}.  Equation~\ref{eq:hydro}
then becomes:
\begin{equation}
  \label{eq:hydro2}
  M(<r) = 1.13 \times 10^{14} \beta
  \frac{T_{\rm x}}{{\rm keV}}
  \frac{r}{{\rm Mpc}}
  \frac{(r/r_{\rm c})^2}{1+(r/r_{\rm c})^2}  M_{\odot} 
\end{equation}
assuming $\mu = 0.59$. The $\beta$-model mass can be written in terms
of density contrast (Equation~\ref{eq:over}) as:
\begin{equation}
  \label{eq:beta2}
  M(\delta,\beta,T_{\rm x}) = 1.1 \times 10^{15} \delta^{-1/2} \beta^{3/2} 
  \left( \frac{T_{\rm x}}{{\rm keV}} \right)^{3/2} 
  \left(1 - 0.01 \frac{\delta r_{c}^2}{\beta T_{\rm x}} \right)^{3/2}
  h^{-1} M_{\odot} 
\end{equation}

We derive $\beta$-model masses using the data of \cite{fukazawa97}
(hereafter F97).  In his study of the metal abundances and enrichment
in the ICM, F97 presents a catalog of 38 clusters with temperatures,
core radii, and $\beta$ parameters derived from ASCA data (see
Table~\ref{tab:clusters} columns 2, 7, and 8).  F97 estimated the
$\beta$ parameter from the ASCA GIS data using a Monte-Carlo method to
take into account the spatial and energy dependence of the GIS PSF and
estimated temperatures by excluding the central region of the x-ray
emission to minimize cooling flow biases.

\begin{figure}
  \psfig{figure=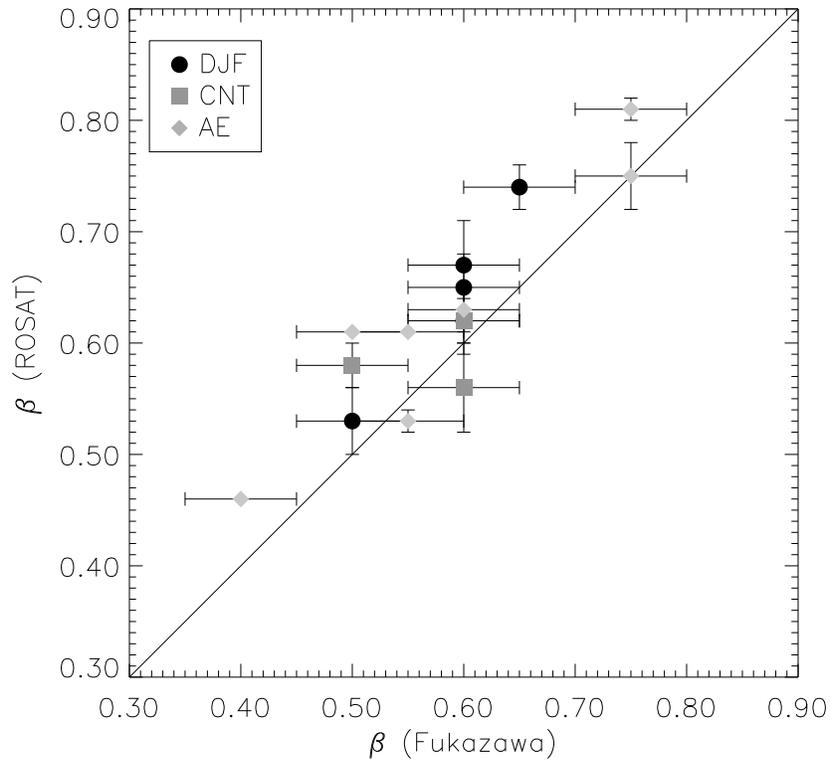}\caption{Comparison F97 ASCA GIS $\beta$
    values to ROSAT $\beta$ values.  The circles are taken from
    \protect\cite{djf95}. The squares are from \protect\cite{cnt97},
    and the diamonds are taken from references listed in
    \protect\cite{ae98}.
  \label{f7.eps}}
\end{figure}

\begin{figure}
  \psfig{figure=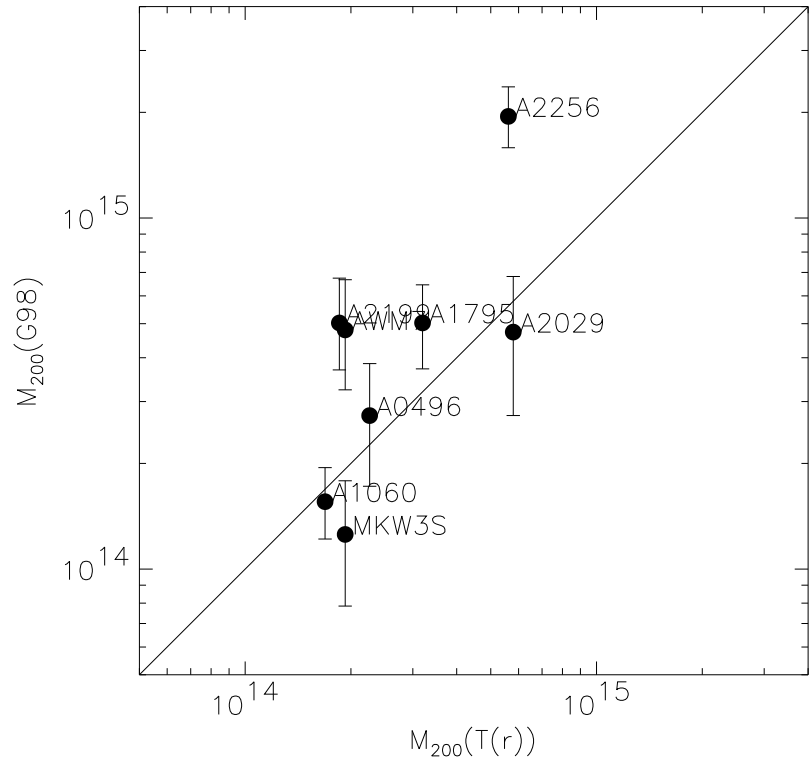}\caption{Comparison of viral and
    temperature resolved mass estimates.
  \label{f8.eps}}
\end{figure}

The F97 data has the advantage of being a homogeneous sample, although
ASCA is not the best instrument for surface brightness fitting due to
its complicated PSF\@. However, the ASCA GIS $\beta$-model fits are
rather insensitive to the presence of cooling flows due to the high
bandpass of the GIS\@.  The clusters are also at low redshifts and so
subtend a large area of the detector minimizing the effect of the poor
resolution of the GIS\@.  We compared the F97 $\beta$ values with
those of derived from ROSAT studies (primarily using the PSPC) of
\cite{djf95}, \cite{cnt97}, and various others taken from the
literature via the compilation of \cite{ae98} (see
Figure~\ref{f7.eps} and Table~\ref{tab:clusters} column 9).  In
general, they agree fairly well although ROSAT $\beta$ values are
higher by an average of about 5\%.

In Figure~\ref{f5.eps}, we plot the estimated $\beta$-model mass at
$\delta = 200$ (using Equation~\ref{eq:beta2}) versus x-ray
temperature.  The relationship ($M \propto T_{\rm x}^{1.78 \pm
  0.05}$) is steeper than that seen using the EMN simulations, virial
masses, or temperature profile masses.  In addition, the relative
normalization with respect to the other mass estimators is a function
of density contrast.  Increasing the density contrast shifts the
$\beta$-model masses lower with respect to the EMN (or virial mass)
relation and closer to the temperature profile masses while decreasing
the density contrast has the opposite effect.

The isothermal $\beta$-model implicitly assumes a dark matter density
profile of $\rho(r) \propto r^{-2}$ in the outer parts of clusters.
The density profiles of clusters in the EMN simulations and clusters
with measured temperature profiles generally seem to follow a steeper
profile $\rho(r) \propto r^{-2.4}$ (see Section~\ref{sec:noniso} and
Figure~\ref{f6.eps}).  Therefore, the dark matter density profile
implied by the isothermal $\beta$ model is probably not an accurate
representation of clusters, and, despite the agreement of the
$\beta$-profile with the observed gas density distribution, the
different dark matter density profiles means that agreement between
the isothermal $\beta$ model estimate and other mass estimates will be
a function of radius (hence density contrast).  The $\beta$-model
masses underestimate the cluster mass at small radii and overestimate
it at large radii.  This has been pointed out previously by
\cite{bs96} and \cite{mfsv98}.  The different density profiles
explain the relative steepness of the $M_{\beta}-T_{\rm x}$
relationship. For an individual cluster, $M_{\beta}/M(T(r)) \propto
r^{0.4}$.  At any given density contrast, $r \propto T_{\rm x}^{1/2}$
and $M(T(r)) \propto T_{\rm x}^{3/2}$, so the $\beta$ model mass
within that overdensity would be $\propto T_{\rm x}^{1.7}$ similar to
the fit given above.

\subsubsection{Gas Density Profiles}

\begin{figure}
  \psfig{figure=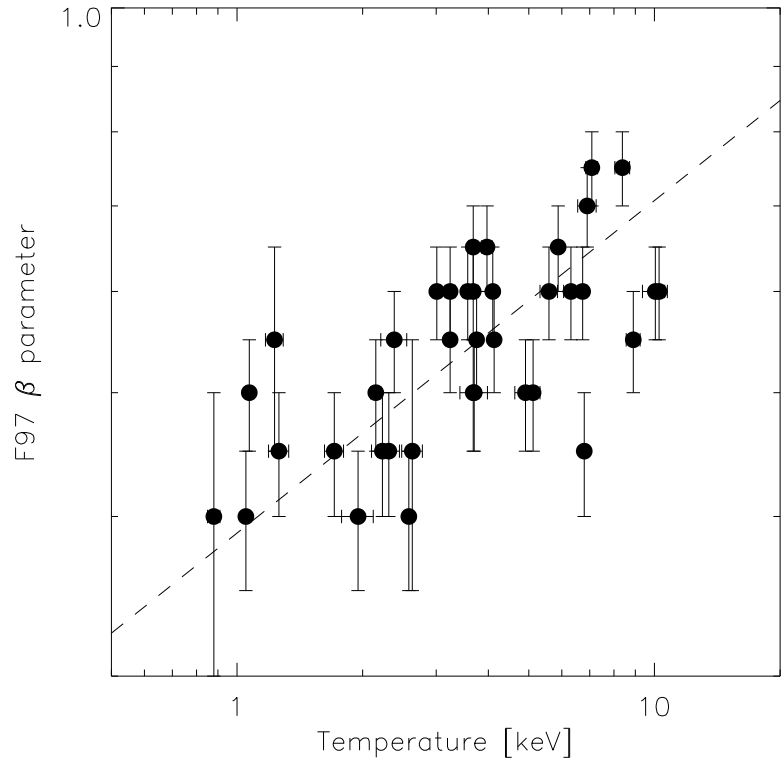}\caption{Relation between $\beta$ and
    $T_{\rm x}$ for F97 data.  The dashed line is a fit to the data
    ($\beta \propto T^{0.26\pm 0.03}$).
  \label{f9.eps}}
\end{figure}

Even if the underlying assumptions of the isothermal $\beta$-model are
flawed, Equation~\ref{eq:bmodel} should still provide a good fit to
the gas density distribution (see \cite{ssm98} for a detailed
discussion).  The EMN relation is basically a $\beta$-model with
$\beta = 0.68$ for all clusters (depending on density contrast)
regardless of x-ray temperature.  However, the F97 data show a
definite correlation of $\beta$ with $T_{\rm x}$ (see
Figure~\ref{f9.eps}).  This indicates that the gas profile becomes
shallower at lower masses.  The relationship closely resembles that of
the \cite{mte97} models with galactic winds, although with a
normalization 20\% lower.

The variation of $\beta$ with $T_{\rm x}$ is unlikely to be an
artifact of the F97 fitting procedure.  As we noted in the previous
section the F97 $\beta$ values generally agree with ROSAT values.  The
correlation of $\beta$ and $T_{\rm x}$ has also been noted previously
in Einstein data \cite{dfj91}, and, more recently, \cite{me97} have
found a similar trend for $\beta$ defined in a non-parametric and
non-azimuthally averaged fashion using PSPC data.  \cite{ae98} also
note the behavior of $\beta$ with $T_{\rm x}$ and the discrepancy
between the $\beta$-model and the expected EMN masses in their sample
of clusters compiled from the literature. In fact, redoing the
proceeding analysis with their sample gives virtually identical
results.


\subsection{Surface Brightness Deprojection}\label{sec:deproj}

Another method for determining the mass of a cluster is x-ray image
deprojection.  The constraint of the observed surface brightness
profile means that the profiles for the variables in
Equation~\ref{eq:deproj1}~\&~\ref{eq:deproj2} can be determined by
specifying one of them.  The usual procedure is to divide the surface
brightness emission into annuli.  The outer pressure must be set in
the outermost annuli (assumed to be due to gas not detected because
its surface brightness is too low).  The observed emissivity in the
outer shell determines the temperature and hence the density.  This
procedure is then stepped inward and repeated.  For more detailed
discussion see \cite{arnaud88}, White et~al.  (hereafter WJF) and
references therein.

WJF present an analysis of 207 clusters using an x-ray image
deprojection analysis of Einstein IPC and HRI data to estimate
the masses of clusters.  WJF choose the functional form of the
gravitational potential as two isothermal spheres, representing the
central galaxy and general cluster potentials.  These are
parameterized by a velocity dispersion and core radius.  For the
central galaxy, these are fixed at 350 km s$^{-1}$ and 1 h$^{-1}$ kpc.
The velocity dispersion of the cluster potential is taken from the
literature or interpolated from the x-ray temperature or luminosity
using an empirical relation.  The core radius is a free parameter in
the analysis which, with the outer pressure, is constrained to produce
a flat temperature profile.  Therefore, the derived gravitational mass
depends on the velocity dispersion, x-ray surface brightness
distribution, and temperature.

WJF determine the mass within the radius that they have x-ray data,
which may be fairly small, while the EMN relation is only valid in the
outer parts of the cluster (i.e.\ low values of density contrast).  As
with the G98 clusters, we can rescale the mass to a $\delta = 200$, at
least for the WJF clusters which have data in the outer parts of the
cluster.  We have rescaled the clusters with data at density contrasts
($= M_{WJF}/(4/3 \pi r_{WJF}^{3} \rho_{c})$) less than 2000 to $\delta
= 200$ assuming $\rho(r) \propto r^{-2}$ as is the case for an
isothermal sphere.

\begin{figure}
  \psfig{figure=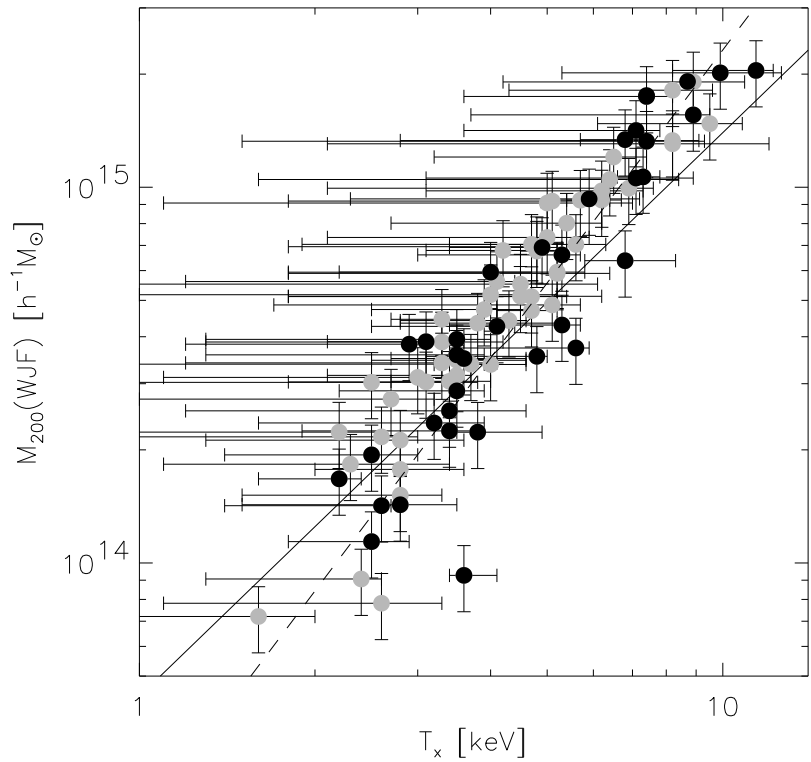}\caption{Mass--Temperature relation for
    WJF clusters using clusters with data for $\delta \leq 2000$.  The
    solid circles are clusters which did not have any interpolated
    input data.  The solid line is the theoretical EMN relation.  The
    dashed line is a fit to the data.  The error bars represent the
    90\% confidence intervals.
  \label{f10.eps}}
\end{figure}

In Figure~\ref{f10.eps}, we show the resulting $M_{200}$ versus
$T_{\rm x}$ relation.  Although the lower $T_{\rm x}$ error bars are
quite large, allowing the the points to be statistically consistent
with the EMN relation, the relation is obviously steeper.  Fitting
this relation using the BCES method gives a $M \propto T_{\rm x}^{2.06
  \pm 0.10}$ somewhat steeper than than found from the F97 data.  The
scatter in mass around this relation is also larger, about 50\%.

Although WJF interpolated $\sigma$ or $T_{\rm x}$ for many clusters,
using only the 45 clusters with data at $\delta < 2000$ which did not
have interpolated data does not alter our results.  Also, the results
reported using a subsample of 19 clusters with better determined
parameters by \cite{wf95} are consistent with the WJF results.
Furthermore, a recent deprojection analysis by \cite{peres98} of 45
clusters with ROSAT PSPC or HRI data gives a similar relationship as
the WJF data, although the Peres et~al.\ data are generally at higher
density contrasts ($\gtrsim 3000$) making the extrapolation to $\delta
= 200$ even more uncertain.

Interpretation of the deprojection results is difficult as the derived
gravitational masses are a combination of optical (the velocity
dispersion) and x-ray data (the core radius and temperature).
However, like the $\beta$-model masses the WJF masses follow a steeper
relation due to the assumption of a potential which has a density
profile which behaves as $\rho(r) \propto r^{-2}$.  The large scatter
is probably due to the use of the velocity dispersion to set the depth
of the potential well.


\section{Conclusions}\label{sec:conclusions}

We have examined the relationship between various galaxy cluster mass
estimators and x-ray gas temperature.  The resulting relationships
generally agree to within $\lesssim 40$\% in mass but with systematic
offsets between different types of mass estimators.  Using G98 virial
masses and ASCA temperatures, we find good agreement with the EMN M--T
relation after removing a few outliers.  X-ray mass estimates using
spatially resolved temperature profiles scale similarly but with a
normalization about 40\% lower.  We note that the \cite{hov98} lensing
estimate mentioned at the end of Section~\ref{sec:mxtx} lies between
these two estimates.

Mass estimates based on the assumption of isothermality like the
$\beta$-model follow a steeper scaling relation due to the implicit
assumption that the dark matter density scales $\rho(r) \propto
r^{-2}$ at large radii while observational and numerical evidence
suggests that clusters follow steeper profiles (i.e.\ $\rho(r) \propto
r^{-2.4}$).  As a consequence, the $\beta$ model underestimates the
mass at low radii and overestimates it at large radii.

The intrinsic dispersion in the true M--T relationship is probably
quite small.  The scatter in the virial mass -- temperature relation
is consistent with most of the scatter being due to the dispersion in
the virial mass estimator.  The small scatter of the masses of
clusters with spatially resolved temperature profiles also indicates
that the dispersion in the M--T relation is probably $\lesssim$ 10\%.
More insight may be gained through simulations of projection effects
on various mass estimators, but the most comprehensive study of these
effects \cite{cen97} is limited to fairly low masses clusters
($\lesssim 4 \times 10^{14}h^{-1} M_{\odot}$ corresponding to $T_{\rm
  x} \lesssim 4$ keV).  Simulations covering a larger mass range are
necessary to better constrain the scatter in the M--T relation.

Our results also show that galaxy clusters seem to be affected by
nongravitational processes such as energy injection by galactic winds.
We find that $\beta$, the asymptotic slope of the gas density profile,
is a function of temperature.  The dependence is consistent with
models of energy injection by galactic winds which tend to make the
gas profile shallower in less massive clusters \cite{me97}.  The
steepening of the $\sigma$--$T_{\rm x}$ relation is also consistent
with energy injection by galactic winds (Bird et~al. 1995).  The good
agreement between the virial and temperature resolved mass estimates
and a $T_{\rm x}^{3/2}$ scaling law indicates that any temperature
changes ($\Delta T$) due to energy injection must be small.  However,
meaningful constraints on $\Delta T$ are difficult given the
relatively few clusters in this regime.

In the future, more optical virial mass estimates for cooler clusters
and larger samples of clusters with resolved temperature profiles will
enable better constraints on the M--T relation.  Better x-ray data
will allow the effect of energy injection on the apparent temperature
and the spatial distribution of the ICM to be disentangled.  Future
x-ray observations (e.g., with AXAF) will produce both the gas density
and temperature maps of clusters and allow maps of the gas entropy to
be constructed which should produce new constraints on energy
injection histories and entropy variations within the cluster
population.  Such entropy maps can then be compared with those
produced from simulations of various heating mechanisms.

\acknowledgements
We would like to that D. White for provide machine readable tables
from the WJF paper and Y. Fukazawa for providing a copy of his
thesis. 


\clearpage
 
\clearpage

\begin{deluxetable}{lccl}
  \tablewidth{0pt}
  \tablecaption{Fitting Results $M_{200} = c_{0} \times 10^{13}\  
    T_{\rm x}^{c_{1}}
  \ h^{-1} M_{\odot}$\label{tab:results} }
  \tablehead{\colhead{Sample} &\colhead{$c_{0}$} & \colhead{$c_{1}$} &
    \colhead{Comments}}
  \startdata
  EMN & $4.42 \pm 0.56$ & 1.5 & \nl
  G98 & $2.94 \pm 0.80$ & $1.71 \pm 0.20$ & all clusters  \nl
  G98 & $3.97 \pm 0.76$ & $1.53 \pm 0.13$ & excluding outliers \nl
  $M(T_{\rm x}(r))$ & $2.72 \pm 0.47$ & $1.48 \pm 0.12$ & \nl
  F97 & $2.18 \pm 0.14$ & $1.78 \pm 0.05$ & \nl
  WJF & $2.02 \pm 0.29$ & $2.06 \pm 0.10$ &  \nl
  \enddata
\end{deluxetable}

\begin{deluxetable}{rrcrcrcccl}
  \small
  \tablewidth{0pt}
  \tablecaption{Cluster Data \label{tab:clusters}} 
  \tablehead{   
    \colhead{Name}  &
    \colhead{$T_{\rm x}$} &
    \colhead{$r_{\rm vir}$} &
    \colhead{$M_{\rm vir}$}&
    \colhead{$r_{\rm x}$} &
    \colhead{$M_{\rm x}$}&
    \colhead{$r_c$}&
    \colhead{$\beta$}  &
    \colhead{$\beta_{ROSAT}$} &
    \colhead{Ref.} \\
    \colhead{}  &
    \colhead{keV} &
    \colhead{$h^{-1}$ Mpc} &
    \colhead{10$^{14} h^{-1} M_{\odot}$}&
    \colhead{$h^{-1}$ Mpc} &
    \colhead{10$^{14} h^{-1} M_{\odot}$}&
    \colhead{$h^{-1}$ Mpc}&
    \colhead{} &
    \colhead{} &
    \colhead{} \\
\colhead{(1)} & 
\colhead{(2)} & 
\colhead{(3)} & 
\colhead{(4)} & 
\colhead{(5)} & 
\colhead{(6)} & 
\colhead{(7)} & 
\colhead{(8)} & 
\colhead{(9)} & 
\colhead{(10)} 
}
  \startdata
2A0335+096& 3.01$\pm$0.07&  \nodata&                  \nodata&     0.50&     1.10&    0.023&      0.60& \nodata&          3\nl
     A0085& 6.31$\pm$0.25&     1.94&   9.88$_{-1.68}^{+2.25}$&  \nodata&  \nodata&    0.086&      0.60& 0.62$\pm$0.03&   10\nl
     A0119& 5.59$\pm$0.27&     1.36&   2.50$_{-0.74}^{+0.90}$&  \nodata&  \nodata&    0.231&      0.60& 0.56$\pm$0.04&   11\nl
     A0194& 2.63$\pm$0.15&  \nodata&                  \nodata&  \nodata&  \nodata&    0.069&      0.45& \nodata&           \nl
     A0262& 2.15$\pm$0.06&     1.05&   1.32$_{-0.25}^{+0.30}$&  \nodata&  \nodata&    0.032&      0.50& 0.53$\pm$0.03&   10\nl
     A0399& 7.00$\pm$0.40&     2.23&  13.45$_{-2.63}^{+2.74}$&  \nodata&  \nodata&  \nodata&   \nodata& \nodata&          1\nl
     A0400& 2.31$\pm$0.14&     1.20&   2.49$_{-0.62}^{+0.73}$&  \nodata&  \nodata&    0.051&      0.45& \nodata&           \nl
     A0401& 8.00$\pm$0.40&     2.30&  13.69$_{-2.50}^{+2.76}$&  \nodata&  \nodata&  \nodata&   \nodata& \nodata&          1\nl
     A0426& 6.79$\pm$0.12&     2.05&   9.08$_{-1.52}^{+2.13}$&  \nodata&  \nodata&    0.020&      0.45& \nodata&           \nl
     A0478& 6.90$\pm$0.35&  \nodata&                  \nodata&  \nodata&  \nodata&    0.077&      0.70& \nodata&           \nl
     A0496& 4.13$\pm$0.08&     1.37&   3.20$_{-0.95}^{+1.04}$&     0.50&     1.50&    0.035&      0.55& \nodata&          4\nl
     A0539& 3.24$\pm$0.09&     1.26&   2.01$_{-0.42}^{+0.52}$&  \nodata&  \nodata&    0.082&      0.60& 0.65$\pm$0.03&   10\nl
     A0754& 9.50$\pm$0.60&     1.32&   4.23$_{-0.72}^{+1.04}$&  \nodata&  \nodata&  \nodata&   \nodata& \nodata&          1\nl
     A1060& 3.24$\pm$0.06&     1.22&   1.90$_{-0.33}^{+0.38}$&     1.00&     1.80&    0.040&      0.55& 0.61&           5,9\nl
     A1656& 8.38$\pm$0.34&  \nodata&   4.97$_{-0.57}^{+0.68}$&  \nodata&  \nodata&    0.208&      0.75& 0.75$\pm$0.03&    9\nl
     A1795& 5.88$\pm$0.14&     1.67&   5.86$_{-1.22}^{+1.33}$&     1.00&     3.00&    0.068&      0.65& 0.74$\pm$0.02& 4,10\nl
     A2029& 9.10$\pm$1.00&     2.33&   6.82$_{-2.30}^{+2.40}$&     0.96&     4.71&  \nodata&   \nodata& \nodata&        1,7\nl
     A2063& 3.68$\pm$0.11&     1.33&   3.04$_{-0.48}^{+0.59}$&  \nodata&  \nodata&    0.067&      0.60& 0.67$\pm$0.04&   10\nl
     A2107& 3.78$\pm$0.19&     1.24&   2.62$_{-0.77}^{+0.81}$&  \nodata&  \nodata&  \nodata&   \nodata& \nodata&          2\nl
     A2142& 9.70$\pm$1.30&     2.26&  17.84$_{-3.49}^{+3.99}$&  \nodata&  \nodata&  \nodata&   \nodata& \nodata&          1\nl
     A2147& 4.91$\pm$0.28&  \nodata&                  \nodata&  \nodata&  \nodata&    0.109&      0.50& \nodata&           \nl
     A2163&13.83$\pm$0.76&  \nodata&                  \nodata&     1.00&    10.70&  \nodata&   \nodata& \nodata&          8\nl
     A2199& 4.10$\pm$0.08&     1.60&   5.71$_{-1.21}^{+1.56}$&     0.50&     1.28&    0.040&      0.60& 0.62$\pm$0.02& 4,11\nl
     A2256& 7.08$\pm$0.23&     2.70&  23.12$_{-3.43}^{+3.95}$&     1.50&     6.00&    0.228&      0.75& 0.81$\pm$0.01&  6,9\nl
     A2319& 8.90$\pm$0.34&     3.09&  39.54$_{-5.12}^{+5.85}$&  \nodata&  \nodata&    0.135&      0.55& \nodata&           \nl
     A2634& 3.70$\pm$0.28&     1.40&   4.31$_{-0.98}^{+1.35}$&  \nodata&  \nodata&    0.123&      0.50& 0.58$\pm$0.02&   11\nl
     A2670& 4.45$\pm$0.20&     1.70&   5.56$_{-0.58}^{+0.72}$&  \nodata&  \nodata&  \nodata&   \nodata& \nodata&          2\nl
     A3266& 8.00$\pm$0.50&     2.21&  11.70$_{-1.65}^{+1.96}$&  \nodata&  \nodata&  \nodata&   \nodata& \nodata&          1\nl
     A3376& 4.00$\pm$0.40&     1.38&   3.64$_{-0.68}^{+0.78}$&  \nodata&  \nodata&  \nodata&   \nodata& \nodata&          1\nl
     A3391& 5.40$\pm$0.60&     1.33&   3.61$_{-1.35}^{+2.20}$&  \nodata&  \nodata&  \nodata&   \nodata& \nodata&          1\nl
     A3395& 5.00$\pm$0.30&     1.70&   5.69$_{-1.33}^{+1.58}$&  \nodata&  \nodata&  \nodata&   \nodata& \nodata&          1\nl
     A3526& 3.68$\pm$0.06&  \nodata&                  \nodata&     0.50&     0.80&    0.038&      0.50& \nodata&          3\nl
     A3558& 5.12$\pm$0.20&     1.95&  11.54$_{-0.91}^{+1.02}$&  \nodata&  \nodata&    0.075&      0.50& 0.61&             9\nl
     A3571& 6.73$\pm$0.17&     2.09&   8.17$_{-2.19}^{+2.40}$&  \nodata&  \nodata&    0.086&      0.60& \nodata&           \nl
     A3667& 7.00$\pm$0.60&     1.94&  11.75$_{-1.26}^{+1.60}$&  \nodata&  \nodata&  \nodata&   \nodata& \nodata&          1\nl
     A4059& 3.97$\pm$0.12&  \nodata&                  \nodata&     0.50&     1.50&    0.075&      0.65& \nodata&          3\nl
      AWM4& 2.38$\pm$0.17&  \nodata&                  \nodata&  \nodata&  \nodata&    0.034&      0.55& 0.47&              \nl
      AWM7& 3.75$\pm$0.09&     1.73&   5.77$_{-1.50}^{+1.80}$&     1.00&     2.00&    0.062&      0.55& 0.53$\pm$0.01&  5,9\nl
     HCG51& 1.23$\pm$0.06&  \nodata&                  \nodata&  \nodata&  \nodata&    0.058&      0.55& \nodata&           \nl
     HCG62& 1.05$\pm$0.02&  \nodata&                  \nodata&  \nodata&  \nodata&    0.003&      0.40& \nodata&           \nl
   HYDRA-A& 3.57$\pm$0.10&  \nodata&                  \nodata&  \nodata&  \nodata&    0.036&      0.60& \nodata&           \nl
     MKW3S& 3.68$\pm$0.09&  \nodata&                  \nodata&     1.00&     2.00&    0.047&      0.65& \nodata&          3\nl
      MKW4& 1.71$\pm$0.09&     1.05&   1.15$_{-0.26}^{+0.35}$&  \nodata&  \nodata&    0.009&      0.45& \nodata&           \nl
     MKW4S& 1.95$\pm$0.17&  \nodata&                  \nodata&  \nodata&  \nodata&    0.023&      0.40& \nodata&           \nl
      MKW9& 2.23$\pm$0.13&  \nodata&                  \nodata&  \nodata&  \nodata&    0.025&      0.45& \nodata&           \nl
   NGC2300& 0.88$\pm$0.03&  \nodata&                  \nodata&  \nodata&  \nodata&    0.024&      0.40& \nodata&           \nl
   NGC5044& 1.07$\pm$0.01&  \nodata&                  \nodata&  \nodata&  \nodata&    0.009&      0.50& \nodata&           \nl
    NGC507& 1.26$\pm$0.07&  \nodata&                  \nodata&  \nodata&  \nodata&    0.014&      0.45& \nodata&           \nl
 OPHIUCHUS&10.26$\pm$0.32&  \nodata&                  \nodata&  \nodata&  \nodata&    0.113&      0.60& \nodata&           \nl
      S753& 2.23$\pm$0.12&     1.07&   1.31$_{-0.50}^{+0.67}$&  \nodata&  \nodata&  \nodata&   \nodata& \nodata&          2\nl
   TRIAUST&10.05$\pm$0.69&  \nodata&                  \nodata&  \nodata&  \nodata&    0.126&      0.60& 0.63$\pm$0.02&    9\nl
     VIRGO& 2.58$\pm$0.03&     1.26&   2.04$_{-0.21}^{+0.28}$&  \nodata&  \nodata&    0.007&      0.40& 0.46&             9\nl
  \tablerefs{
(1) $T_{\rm x}$ from Markevitch (1998) 
(2) $T_{\rm x}$ derived for this paper
(3) $M_{\rm x}$ \& $r_{\rm x}$ from Ohashi (1997)
(4) $M_{\rm x}$ \& $r_{\rm x}$ from Mushotzky et al. (1994)
(5) $M_{\rm x}$ \& $r_{\rm x}$ from Loewenstein \& Mushotzky (1996)  
(6) $M_{\rm x}$ \& $r_{\rm x}$ from Markevitch \& Vikhlinin (1997) 
(7) $M_{\rm x}$ \& $r_{\rm x}$ from Sarazin, Wise, \& Markevitch
  (1998), their value of $T_{\rm x} =8.69_{-0.36}^{+0.28}$ is used
  in Section 4.1
(8) $M_{\rm x}$ \& $r_{\rm x}$ from Markevitch et al. (1996) 
(9) $\beta_{ROSAT}$ from Arnaud \& Evrard (1998)
(10) $\beta_{ROSAT}$ from David, Jones, \& Forman (1995)
(11) $\beta_{ROSAT}$ from Cirimele, Nesci, \& Trevese (1997)
 }
\enddata
\end{deluxetable}

\end{document}